# Coherently amplified ultrafast imaging using a free-electron interferometer

Tomer Bucher[1*], Harel Nahari[1*], Hanan Herzig Sheinfux[2*], Ron Ruimy[1*], Arthur Niedermayr[1], Raphael Dahan[1], Qinghui Yan[1], Yuval Adiv[1], Michael Yannai[1], Jialin Chen[1], Yaniv Kurman[1], Sang Tae Park[3], Daniel J. Masiel[3], Eli Janzen[4], James H. Edgar[4], Fabrizio Carbone[5], Guy Bartal[1], Shai Tsesses[1], Frank H.L. Koppens[2,6], Giovanni Maria Vanacore[7], Ido Kaminer[1]

1. Andrea & Erna Viterbi Department of Electrical and Computer Engineering, Technion–Israel Institute of Technology, 3200003 Haifa, Israel.
2. ICFO-Institut de Ciències Fotòniques, The Barcelona Institute of Science and Technology, Av. Carl Friedrich Gauss 3, 08860 Castelldefels (Barcelona), Spain.
3. Integrated Dynamic Electron Solutions, Inc. (a JEOL company), Pleasanton, CA 94588, USA.
4. Tim Taylor Department of Chemical Engineering, Kansas State University, Manhattan, KS 66506, USA.
5. Laboratory for Ultrafast Microscopy and Electron Scattering (LUMES), École Polytechnique Fédérale de Lausanne, 1015 Lausanne, Switzerland.
6. ICREA-Institució Catalana de Recerca i Estudis Avanats, Passeig Lluís Companys 23, 08010 Barcelona, Spain.
7. Department of Materials Science, University of Milano-Bicocca, Milano 20126, Italy.

lightmatterinteractions@gmail.com; kaminer@technion.ac.il

**Accessing the low-energy non-equilibrium dynamics of materials and their polaritons with simultaneous high spatial and temporal resolution has been a bold frontier of electron microscopy in recent years. One of the main challenges lies in the ability to retrieve extremely weak signals while simultaneously disentangling amplitude and phase information. Here, we present Free-Electron Ramsey Imaging (FERI), a microscopy approach based on light-induced electron modulation that enables coherent amplification of optical near-fields in electron imaging. We provide simultaneous time-, space-, and phase-resolved measurements of a micro-drum made from a hexagonal boron nitride membrane, visualizing the sub-cycle dynamics of 2D polariton wavepackets therein. The phase-resolved measurement reveals vortex–anti-vortex singularities on the polariton wavefronts, together with an intriguing phenomenon of a traveling wave mimicking the amplitude profile of a standing wave. Our experiments show a 20-fold coherent amplification of the near-field signal compared to conventional electron near-field imaging, resolving peak field intensities in the order of ~W/cm$^2$, corresponding to field amplitudes of a few kV/m. As a result, our work paves the way for spatio-temporal electron microscopy of biological specimens and quantum materials, exciting yet delicate samples that are currently difficult to investigate.**

### Modulated electrons: from accelerators to microscopy

Free-electron physics has had a profound impact on many areas of science and technology, from electron microscopes and X-ray sources to microwave sources and accelerators. At the heart of all these applications lies the fundamental interaction between free electrons and electromagnetic fields (*1–3*). This interaction can be enhanced by modulating the electrons before (and in some cases after) the interaction (*4*, *5*). This electron modulation is key to applications such as electron radiation and electron acceleration, as recently demonstrated in dielectric laser accelerators (*6*, *7*).

Electron modulation can be achieved through either classical or quantum-mechanical methods, by shaping the longitudinal electron distribution or electron wavepacket, respectively. In both cases, the electron modulation can be performed by laser interaction, as inspired by research into photon-induced near-field electron microscopy (PINEM) (*8*). PINEM was originally conceived as an imaging technique realized in an ultrafast transmission electron microscope (UTEM). Nevertheless, at its core PINEM relies on a fundamental interaction: inelastic scattering of free-electron pulses by optical near-fields. This interaction allows one to reconstruct the near-field amplitudes down to single-nm spatial resolution (*8–11*) and sub-ps temporal resolution (*12*). In fact, PINEM enabled a range of imaging modalities in a variety of nanophotonic and condensed matter systems, including surface polaritons (*9*, *12, 13*), nano-cavities (*10*, *11*) and nanoscale plasma or charge distributions (*14, 15*).

Apart from imaging applications, the electron-laser interaction in PINEM-type experiments has also inspired various electron modulation schemes. This modulation allows one to reconstruct the electron's quantum state (*16*, *17*) and to extract the coherence and decoherence times of quantum emitters (*18*). A notable application of PINEM-modulated electrons is to retrieve the temporal phase information of the optical field in a scheme coined free-electron Ramsey-type phase control (*19*). Such a scheme has been used to reconstruct the sub-cycle dynamics of terahertz (*20*) and optical (*21-24*) fields. However, despite the wide range of applications enabled by electron modulation, this approach has never been used to amplify microscopy itself, i.e., it was not used to increase the sensitivity in near-field imaging with electron microscopes.

Here, we propose and demonstrate coherent amplification of electron imaging of optical near-fields, relying on optically modulated free electrons for probing the investigated sample (Fig. 1a). Conceptually, our imaging scheme can be thought of as a frequency-tunable Ramsey interferometer (*25*) performing measurements at each point in the sample simultaneously. Thus, we dubbed this imaging scheme as Free-Electron Ramsey Imaging (FERI). We use FERI to specifically demonstrate coherently amplified imaging of polariton dynamics in a hexagonal boron nitride (hBN) flake held on a gold frame, forming a micro-drum structure supporting novel phonon-polariton wave excitations (Fig. 1b,c). Compared to conventional PINEM, we show that FERI yields 20-fold coherent amplification of the raw signal contrast, with further enhancement thanks to the electron-field interaction theory that underlies the algorithmic scheme. The overall enhancement enables the retrieval of an image when adopting incident intensities as low as ~W/cm$^2$, paving the way to new kinds of microscopy experiments in scenarios that were previously beyond reach due to electron or laser dose sensitivity. Examples include quantum materials like high-$T_c$ superconductors (*26*), and even soft matter, where maximizing the signal with limited dose is essential.

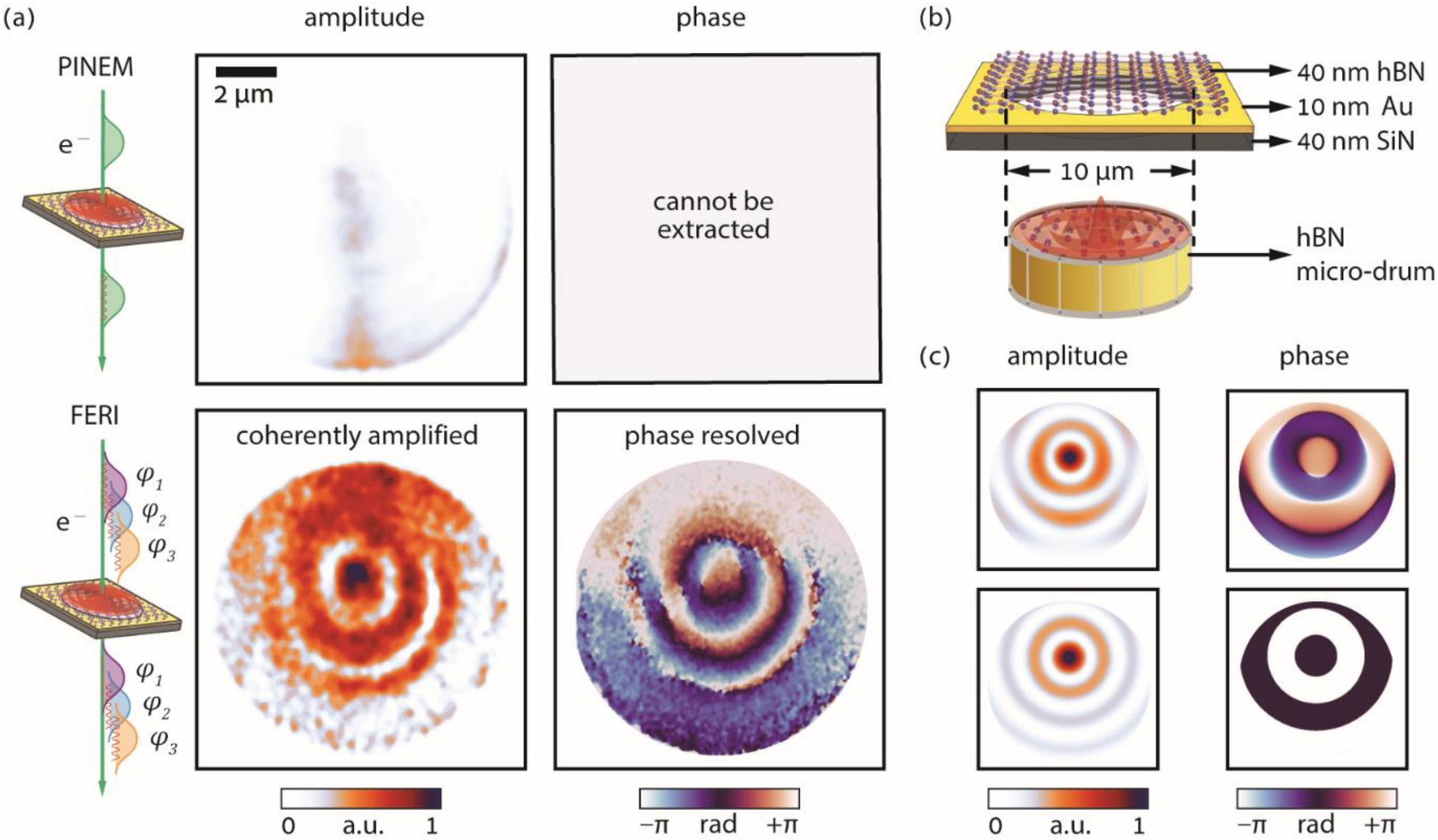

**Fig. 1 | Free-Electron Ramsey Imaging (FERI): utilizing electron modulation for coherently amplified imaging of optical near-fields. (a)** Conventional PINEM (top) scheme compared to FERI (bottom). Modulated electron interference enables the measurement of weaker field intensities and phase resolution, which are inaccessible using conventional PINEM with the same electron dose and laser intensity. The FERI and PINEM amplitude images under the "amplitude" title are normalized to have the same contrast as achieved experimentally (see Methods). **(b)** The platform in which we demonstrate the concepts of this work is a micro-drum made from an hBN membrane. A pump pulse in the mid-IR couples at the circular boundary, exciting a phonon-polariton wavepacket that propagates and interferes, resulting in a wave pattern with an amplitude profile that resembles a standing wave in a drum. **(c)** The simulations show that waves of fundamentally different types can look similar in their amplitude profiles while having drastically different phase profiles. Conventional PINEM extracts the amplitude profile, while FERI additionally extracts the phase profile and can thus distinguish between the different wave types. For example, the amplitude profile in panel (a) resembles a standing-wave mode of a drum. However, the phase profile reveals that the wave is, in fact, not a standing wave, but a different wave phenomenon with a phase resembling that of a traveling wave. The polaritons travel inside a 40 nm-thick layer of hBN, which is held on top of a 10 nm-thick gold frame situated on a 40 nm-thick silicon nitride substrate. A circular hole with a 5 microns radius is drilled trough the gold and silicon nitride. Notice that the illumination angle breaks the up-down symmetry, inducing a progressive phase that shifts the interference pattern from the center.

It should be noted that recent advances in ultrafast electron diffraction and microscopy have enabled imaging of phonon dynamics (*27*). Particularly, ultrafast transmission electron microscopy has been applied to visualize the dynamics of acoustic phonon wave propagation in 2D materials (*27, 28*). Our work takes a completely different path, investigating optical phonons hybridized with electromagnetic fields, also known as phonon-polaritons *(29-31).* In comparison with acoustic phonons, the (optical) phonon polaritons have completely different dispersion relations, e.g., hyperbolic dispersion as in the hBN phonon polaritons that we investigate in this work. The phonon polaritons also evolve on shorter timescales (from a few fs to a few ps) relative to their acoustic counterparts. Such time scales necessitate a method like FERI to investigate the sub-cycle oscillations, i.e., study the complete phase dynamics. The sub-cycle dynamics was impossible to observe without FERI, by just using conventional PINEM (*12*).

**Standing-wave profiles and traveling-wave phases**

Combining the phase- and time-resolved capabilities of FERI reveals a surprising polariton phenomenon. The polaritonic wavepacket (whose amplitude is linear to that of the excitation pump laser) exhibits a multi-ring amplitude profile resembling a standing-wave (Fig. 1c left panels). Indeed, based on amplitude measurements alone (i.e., using conventional PINEM), the polariton could be identified as a standing wave. However, extracting the phase information from FERI reveals that the wavepacket acquires a continuous phase as it propagates, in stark contrast with a standing wave (Fig. 1c right panels). Therefore, the polariton wavepacket cannot be classified as either a conventional traveling wave or as a standing wave. This is a new and intriguing observation in an hBN phonon polariton micro-drum, which is reminiscent of known effects in acoustics (*31*). The observation of such behavior has been made possible only thanks to the phase-resolving capability of our FERI technique and warrants additional future research.

**Experimental system for phase- and time-resolved imaging**

We realize FERI in a modified ultrafast transmission electron microscope (UTEM) (Fig. 2a). In a conventional system, a femtosecond laser pulse is split into two pulses. One pulse (pump) impinges on the investigated sample, and the other pulse (probe) is frequency-up-converted into a UV pulse using fourth harmonic generation (FHG), which photo-emits an electron pulse from the cathode. Together, such a system implements a *pump-probe* scheme with an electron as the probe. In contrast, in our modified system, the pump pulse is split into two pulses, as shown in Fig. 2a. Specifically for the experiments shown here, the pulse is frequency-down-converted to the mid-IR using difference frequency generation (DFG) in an optical parametric amplifier (OPA) before being split. The first pump pulse impinges on the investigated sample, while the second pump pulse does not interact with the sample, and instead modulates the electron before that electron probes the sample (the reference and sample order can be changed with similar outcomes). Together, the modified system implements a *pump-pump-probe* scheme, with the electron as the probe. Variants of this scheme were realized in transmission electron microscopes under different constrains, for example, using separate light-coupling ports for continuous wave operation (*23, 24*) or combined into a single port for pulsed operations, splitting the two points of interaction (*20*), or keeping them together (*13, 32*).

We modify the conventional operation of the ultrafast transmission electron microscope by installing a specially designed component called the *photonic electron modulator* (PELM, see Fig. 5). This modification enables independent tunability of the two pulses in two separate light-coupling ports, controlling their relative delay, intensity, and polarization, as well as other options (described in the methods chapter). Specifically, the electron modulation is implemented by a flat Al-coated $Si_3N_4$ membrane, which is tilted at an angle of 41° with respect to the electron propagation direction to ensure transverse electron-light phase matching. The electrons penetrate through this membrane simultaneously with the laser pulse impinging on it. The flat membrane acts as a light-reflecting mirror, providing the reference interaction in Fig. 2a, while also being relatively electron-transparent to allow further interaction with our micro-drum.

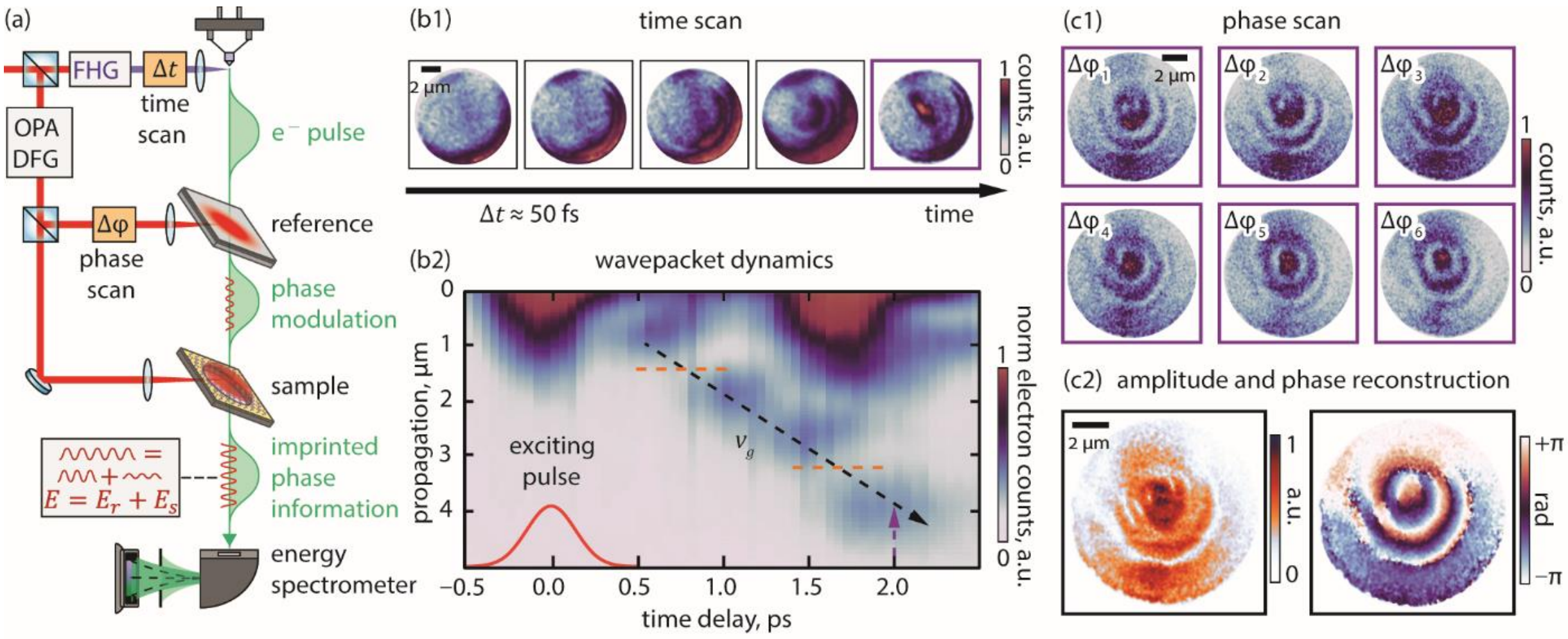

**Fig. 2 | Time- and phase-resolved imaging of polariton wavepackets. (a)** Free-electron Ramsey imaging (FERI) scheme: splitting a femtosecond laser pulse to three parts, enabling both ultrafast time-resolved and phase-resolved imaging. Two delay stages are used to control the timing of the laser pulses with respect to the electron pulse. The effective field modulating the electron after both interactions ($E$) is the superposition of the field at the reference ($E_r$) and the field at the sample ($E_s$) – an interference mediated by the electron. **(b1) Time-resolved imaging** presenting selected raw PINEM data frames at different time delays between the UV pulse and the exciting laser with 50 fs time differences. Full data shown in movies SM1 and SM2. The dark parts correspond to the polaritonic wavepacket propagating from the edges to the center and undergoing interference with other parts of the wavepackets. **(b2)** The polariton wavepacket dynamics is presented in a map where the propagation axis represents the dynamics along the radius, averaged along a semicircle. The polariton wavepacket propagates with a group velocity of $v_g \approx c/200$, marked by a black dashed arrow. The wavepacket "hops" over certain radii at which we observe very low intensity (e.g., see dashed orange lines). This behavior is not expected for conventional traveling waves that propagate continuously, and not expected for conventional standing waves that do not propagate. Instead, this behavior is consistent with the observation (Fig. 1) of a simultaneous standing-wave amplitude and traveling-wave phase profiles. The red feature at the top right of the map is due to the arrival of a second excitation pulse, coupling at the edge. **(c1) Phase-resolved imaging** presenting selected raw FERI data frames at different sub-cycle delays (full data in movie SM3-5), corresponding to different relative phases between the modulation and sample excitation. **(c2)** Algorithm-aided reconstruction of the amplitude and phase profiles from the raw FERI data frames using all phases. The phase-scan is performed around a time delay of 2 ps, marked by the purple frame in (b1) and a purple dashed arrow in (b2).

## Free-electron Ramsey imaging (FERI): concept and results

Our FERI experiment extends on previous works (*13, 20, 23, 24*) by one important development: *coherent amplification of electron imaging*. All the above-mentioned modified ultrafast transmission electron microscopes provide a *controllable phase relation* between the electron modulation and the induced near-field on the investigated sample (this is possible because both pump pulses were split from the same laser pulse). Our demonstration of coherently amplified imaging requires an additional development, which can be summarized in two stages: [1] Acquiring raw modulated-electron images, by uniformly scanning over the relative phase. [2] Applying a reconstruction algorithm on the raw images, extracting high-contrast phase and amplitude of the sample near-field. We calibrate our measurement by acquiring electron energy-filtered images without inducing field on the sample, and by taking separate electron energy spectra measurements to estimate the strength of the reference interaction (see methods chapter).After its interaction with the sample, each modulated electron can be characterized by measuring its electron energy spectrum, which depends on the sample near-field and its phase relative to the reference interaction. We specifically use electron energy filtering on the electron image (methods chapter). This way, for each choice of modulation phase, the raw electron image provides different information about the near-field at the sample. By combining the raw images (methods chapter), our algorithm yields the phase and the amplified contrast in the amplitude profile, meaning that weaker fields can be reconstructed with the same electron dose. For example, Fig. 1a compares two images acquired with the same total electron dose and same field intensity on the sample – showing the advantage of FERI over conventional PINEM, with a 20-fold increase in the resulting signal contrast.

At every pixel, our scheme is analogous to Ramsey interferometry in atomic physics (*33*) or to homodyne detection in optics (*34*). However, this analogy is not precise. Only for special cases, depending on the interaction strength and on the distance between the modulation stage and the sample, our algorithm reproduces the regime where the Radon transform is used in conventional homodyne detection. But for the general case, our algorithm goes beyond the Radon transform (*35*). This more general case is necessary for our experiment.

Next, we apply FERI to investigate phonon-polariton wavepackets in hBN and measure their group and phase dynamics. Fig. 2b illustrates the wavepacket propagation from the Au edge of the micro-drum towards its center (a radial half-circle profile of the movies SM1 and SM2 in the Supplementary Information). The slope of the dashed black arrow in Fig. 2 (b2) corresponds to the group velocity of the excited wavepacket which is measured to be $v_g \approx c/200$. By using modulated electrons and by performing a sub-cycle delay scan between the two points of interaction, the *phase* dynamics of the phonon-polariton wavepacket is retrieved (utilizing an optimization process, methods chapter).

A peculiar feature of the polariton wavepacket dynamics is highlighted in Fig. 2b, where we present the polaritonic field intensity averaged along a semicircle as a function of radius. The laser pulse excites a narrow wavepacket (roughly 1 micron wide) that propagates from the edge toward the center. The black dashed arrow marks the polariton wavepacket propagation, revealing the unusual feature of wavepacket "hopping": Certain rings at fixed radii remain low in their intensity throughout the dynamics. These rings appear as dark lines in Fig. 2b (marked by dashed orange lines). One typically expects stationary features of zero intensity to be a hallmark of standing waves. But this is not the case here, since the polariton wavepacket has a wide bandwidth and its width is too small. Rather, this hopping behavior is consistent with the observation from Fig. 1 about the polariton wavepacket having simultaneously a standing-wave-like amplitude profile and a traveling-wave-like phase accumulation.

**Sub-cycle dynamics of polariton wavepackets**

Fig. 3a depicts the phase profiles for different excitation wavelengths, each retrieved from a scan of sub-cycle time delays on the modulated electrons, as shown in Fig. 2c. These phase profiles provide a direct measurement of the phonon-polariton wavenumber. We extract the wavenumbers from the inverse distance over which the phase of the wavepacket accumulates $2\pi$, and compare them with theory in Fig. 3b (black circles on top of the theoretical dispersion of hBN phonon polaritons). The direct measurement of the dispersion further establishes that the phenomenon we observe is phonon-polariton dynamics.

The phase-reconstruction capability of FERI enables to pinpoint another intriguing phonon-polariton phenomenon. We show, for the first time in PINEM and generally in electron microscopy, the coexistence of vortex-anti-vortex singularities at the nodal point located in the center of the micro-drum wave (Fig. 3c). This occurrence is a consequence of symmetry breaking. The circular symmetry of the geometry is broken by the exciting laser pulse, which comes from a particular direction, splitting the original nodal lines of zero-amplitude and no orbital angular momentum into a pair of polariton vortices of opposite chirality (*36*). The opposite chirality of the vortices guarantees that the left-right mirror symmetry is preserved (as the laser comes from the up-down direction, breaking only this symmetry). We note that vortices have been previously observed in conventional PINEM (*17*), but identifying the vortex chirality is only made possible in FERI.

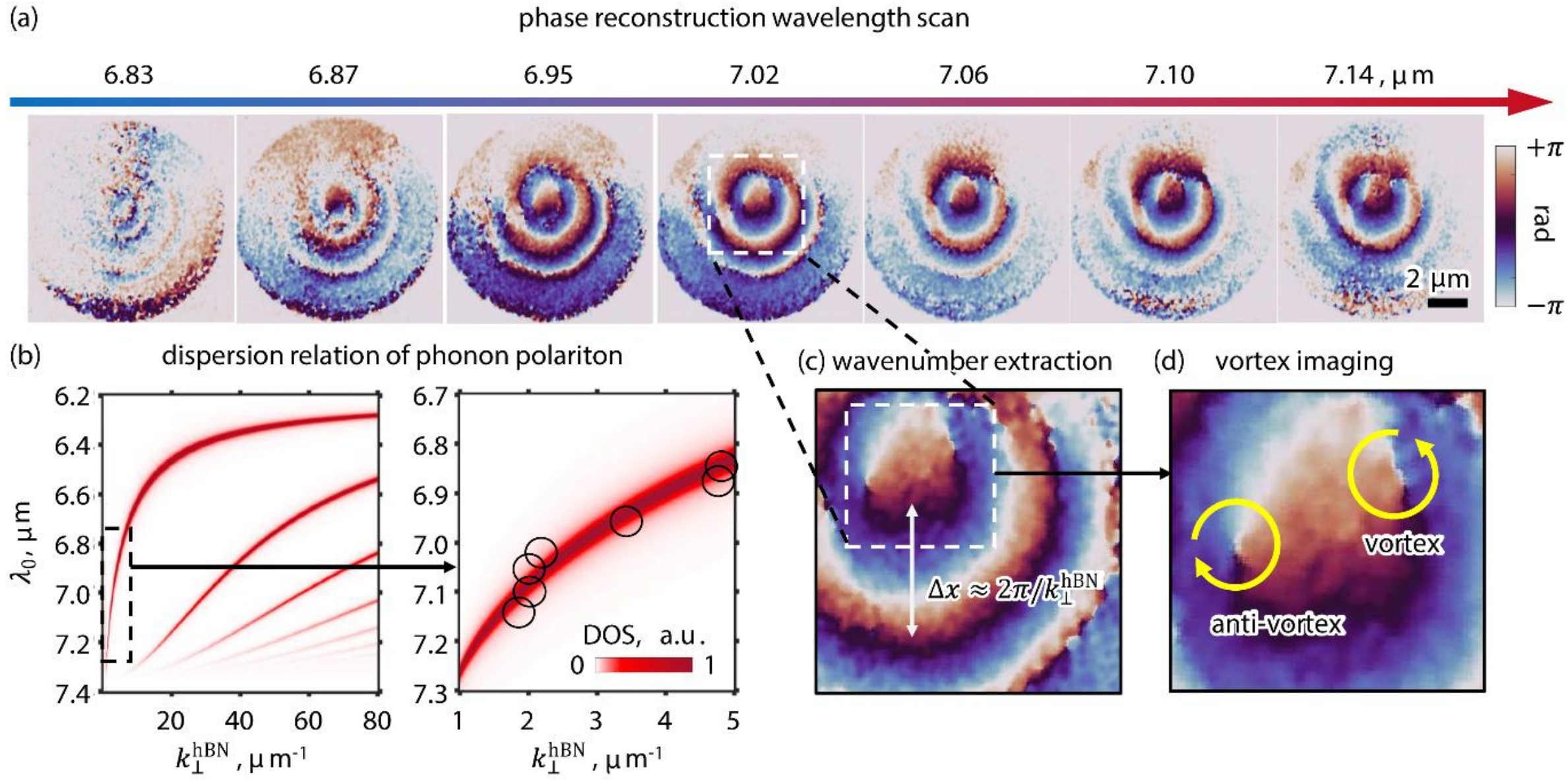

**Fig. 3 | Intrinsic phonon-polariton properties extracted via FERI: direct measurement of the polariton phase velocity and identification of a pair of vortex and anti-vortex. (a)** FERI phase-reconstruction for different wavelengths. The exciting laser power at the presented different wavelengths is around 2 nW. **(b)** We use the distance between consecutive equal-phase circles for direct extraction of the polariton's wavenumber (and phase velocity). The dispersion relation of phonon-polaritons in hBN, with circles marking the extracted wavenumbers. The density of states (DOS) dispersion relation is achieved by calculating the reflection coefficient of the air-hBN-air scattering, where the thickness of the hBN is taken to be 45 nm, as estimated from a TEM measurement. **(c-d)** Zoom-in on a phonon-polariton phase profile portrays that the nodal point at the center of the micro-drum wave is comprised of a vortex-anti-vortex pair, an inevitable result of symmetry breaking in the system (*57*). By extracting the exact phase, FERI enables measuring the chirality of each vortex.

## Coherent amplification of electron imaging contrast

In this section, we quantify the coherent amplification of the electron image contrast (defined as the difference between upper and lower tenth percentile of the signal). The signal is normalized by having the same electron dose on the sample in all measurements. Recalling that the conventional PINEM signal of an electron with velocity $v$ moving along $z$ depends on a dimensionless parameter $g(x,y) = \frac{q_e}{\hbar\omega}\int_{-\infty}^{\infty} dz E_z(x,y,z) e^{-iz\omega/v}$ (*37, 38*), with $q_e$ being the electron charge and $\omega$ the angular laser frequency. We denote by $g_s$ the PINEM parameter for the sample signal field $E_s$, and by $g_r$ the PINEM parameter for the reference field $E_r$. Fig. 4c presents the contrast values extracted from measurements of FERI in different areas of the image that have different field intensities. In each region, we choose the relative phase in FERI such that maximal contrast is achieved. In all cases, the contrast is larger than the contrast in conventional PINEM. Comparison of the raw data from FERI and PINEM in Fig. 4c already shows the enhanced contrast by coherent amplification (despite an acquisition time shorter by a factor of 21). Overall, the contrasts ratios correspond qualitatively to the theory in Fig. 4a. In some areas of the sample, the amplification factor approaches 20, exemplifying the advantage of coherent amplification.

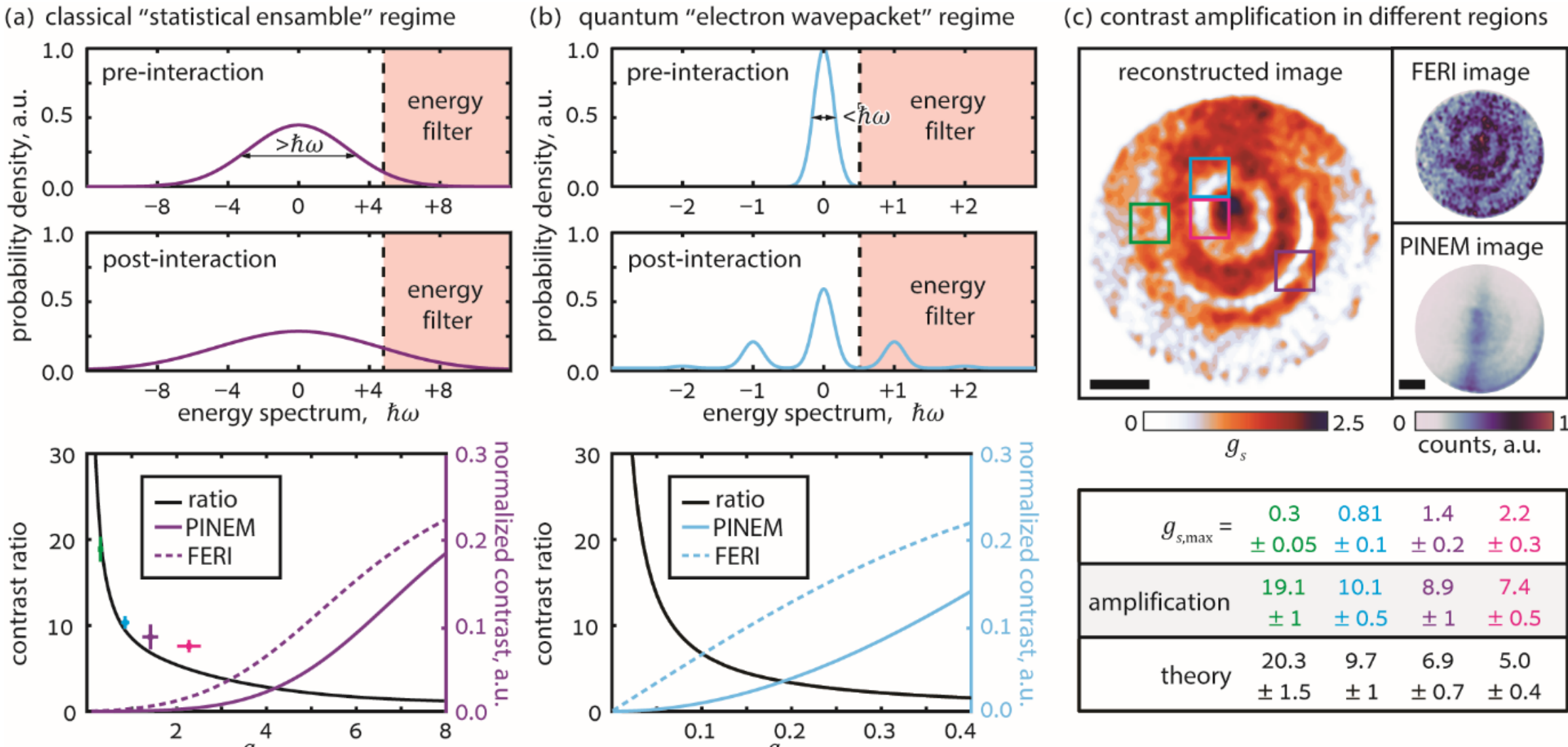

**Fig. 4 | Prospects of FERI for precision measurements of weak fields, and comparison between the classical and quantum regimes. (a)** Coherent amplification in the classical regime (electron energy spread wider than $\hbar\omega$, taken in our simulation to be $3.5\hbar\omega$ corresponding to our experiment), which is demonstrated in this work. The values from the table in (c) are marked on the theoretical curve (error bars correspond to the values in the table). The experimental parameters are used to draw the theory curves. **(b)** Coherent amplification in the quantum regime (electron energy spread narrower than $\hbar\omega$), showing prospects for additional amplification beyond the classical regime, achieving higher contrast for smaller signals at the sample (*35*). **(c)** Coherent amplification of electron imaging, showing the contrast amplification in different regions of the sample. (top left) The regions from which we extract the corresponding values of $g_s$ are marked on the reconstructed image. (bottom) The table presents experimentally measured $g_s$ values with the corresponding extracted and theoretical amplification factors. The error values assigned to $g_{s,\mathrm{max}}$ are defined by the difference between the 90$^{\mathrm{th}}$ percentile to the max value of the region. This derives the error values of the contrast. (top right) We extract the contrast from the regions in the raw data collected by FERI and by conventional PINEM. Although the collection time in FERI (20s) is 21 times shorter than that of the PINEM image (420s), the features of the phonon-polariton near-field are already more visible in the FERI image. The total integration time of 420s corresponds to an electron dose of roughly $0.2\frac{e}{\mathrm{nm}^2}$ (estimated from measuring the electron counts on the camera without the sample). Scale bars are 2 μm.

## Discussion

When applying FERI theory in actual experiments, we separate between two cases: [1] electron energy spread $\Delta E$ larger than the photon energy $\hbar\omega$ (classical regime, Fig. 4a), and [2] $\Delta E$ smaller than $\hbar\omega$ (quantum regime, Fig. 4b). Comparing the two regimes shows the advantage of the quantum regime for small $|g_s|$, whereas the situation is more complex for larger $|g_s|$ as in our experiment. The advantage of the quantum regime lies in the fact that it is easier to distinguish electrons that absorbed or emitted even just one photon from those that did not change their energy. We note that Eq. 4 in the methods chapter generalizes both regimes under a single formula.

To estimate the minimal field amplitude $E_s$ that is possible to image with FERI, consider that the PINEM parameter interaction strength satisfies $g_s \approx \frac{E_s q_e L}{\hbar\omega}$ (*37, 38*), with $L$ being an effective interaction length between the electron and the field. For the phonon polaritons we observed, $L$ is typically a few $\mu$m. We estimate the weaker features that were successfully resolved in Fig. 4 to have $g_s \sim 0.2$ (as extracted from the reconstruction), corresponding to a field $E_s$ of few kV/m or intensities around 1 W/cm$^2$.

It is important to differentiate the phase-resolved near-field imaging capabilities enabled by modulated electrons from the most established methods of phase-resolved near-field measurements. In particular, photo-emission electron microscopy (PEEM) (*39*) is based on the nonlinear response of the target sample, which becomes less efficient for low-energy photons as in the mid-IR. Hence, this approach is not easily applied to investigate phonon polaritons in 2D materials, which are typically present in the mid-IR or at lower photon energies. Moreover, PEEM is limited to probing phenomena near surfaces, unlike FERI, which is inherently sensitive to the bulk. The ability to probe the internal optical field inside the bulk is relevant in scenarios where the fields are strongly confined inside the bulk (more than in our work), such as guided optical modes in semiconductors or polaritons in hyperbolic materials in the optical range. This latter difference also distinguishes FERI from scattering-type scanning nearfield optical microscopy (sSNOM) (*40, 41*). Generally, near-field measurements using scanning tips are well-established over wide spectral bandwidths and can be performed as well with an ultrafast source (*39*). However, the reconstruction of the polaritonic field from the measurement is often limited by requiring precise modelling and assumptions on the tip-field interaction. More generally, unlike tip-based techniques that inevitably disturb the near-field, the electron probe can measure the near-field without altering it.

**Outlook**

The ability of FERI to measure the amplitude and phase of the fields also *inside the bulk* of a sample opens a set of intriguing capabilities. By collecting angle-dependent measurements, FERI can be used for full tomography and extract the 3D profile of an electromagnetic mode. This would be a completely new type of measurement. A notable application that could arise from full tomography is imaging deeply confined polaritons, such as acoustic graphene polaritons (*42*) and picocavity polaritons (*43, 44*). These polaritons are often inaccessible by conventional surface techniques. FERI facilitates the detection of such polaritons also by coherently amplifying their low intensity.

The coherently amplified sensitivity of FERI to low-intensity fields may also provide access to highly desired polariton nonlinearities in van der Waals and phononic materials (*45*). Such nonlinearities are typically challenging to observe due to the small variation in their polaritonic wavelengths. Looking forward, FERI's ability to image low-intensity fields may open new modalities for imaging biological samples and other dose-sensitive materials (*46, 47*), which were so far beyond reach for PINEM because they necessitated field intensities that were too high (*38*). Electron modulation in FERI reduces the necessary field intensity while increasing the amount of information that can be extracted by each electron. Thus, FERI reduces the number of electrons needed to pass by the sample, minimizing both laser damage and electron radiation damage.

Enticing goals that could now be attempted include quantum materials (*48, 49*), such as encapsulated 2D superconductors (*26*) and others that were so far beyond reach. From a fundamental perspective, coherent amplification of the *intrinsic* electron-photon interaction of quantum electrodynamics could eventually enable observations of electron interactions with few-photons states or with squeezed vacuum fluctuations (*50-53*).

## Acknowledgments

This work and especially the development and installation of the PELM are part of the SMART-electron Project that has received funding from the European Union's Horizon 2020 Research and Innovation Programme under Grant Agreement No. 964591. Measurements were conducted in I.K.'s UTEM laboratory in the electron microscopy centre (MIKA) in the Department of Materials Science and Engineering of the Technion. We acknowledge the Hellen Diller Quantum Center for supporting this research. The experimental effort is funded by the Gordon and Betty Moore Foundation, through I.K.'s Grant GBMF11473. A.N. acknowledges funding from the Swiss National Science Foundation (SNSF) through project P500PT_214469. S.T. acknowledges the generous support of the Adams fellowship of the Israeli Academy of Science and Humanities; the Yad Hanadiv foundation through the Rothschild fellowship; the VATAT-Quantum fellowship by the Israel Council for Higher Education; the Helen Diller Quantum Center post-doctoral fellowship; and the Viterbi fellowship of the Technion – Israel Institute of Technology.

## Author Contributions Statement

The sample fabrication was done by H.H.S. The hBN crystals were grown by E.J. and J.H.E. The measurements were performed by H.N., A.N., R.D., Y.A. and M.Y. The theory was developed by R.R., T.B., Q.Y. and J.C. The data analysis was performed by T.B. and R.R. The design and installation of the PELM in the electron microscope was performed by R.D., S.T.P., D.J.M. and G.M.V. The experiment was designed by T.B., H.H.S., H.N., A.N., S.T., Y.A., M.Y., R.D. and I.K. The work was supervised by F.C., G.B., F.H.L.K, G.M.V. and I.K. All authors contributed significantly to the analysis, discussion, and writing of this work.

## Competing Interests Statement

The authors declare no competing interests.

## Methods

### 1. Experimental setup

The measurements were performed in an ultrafast transmission electron microscope (UTEM) based on a JEOL JEM-2100 Plus TEM with a $LaB_6$ electron gun and acceleration voltage of 200 kV (Fig. 5a-b). The UTEM operates as a pump–probe setup driven by a 40 W, 1030 nm, ~270 fs laser (Carbide, Light Conversion) operating at a 1 MHz repetition rate. The output of this laser is split into two pulses: one pulse is upconverted into 266 nm (UV) using two stages of second-harmonic generation and excites the $LaB_6$ cathode, generating single-electron probe pulses at the laser repetition rate. These electron pulses travel down the microscope column, passing the reference and sample interactions (see below), and are eventually measured by one of the installed electron detectors. The second pulse is converted into variable wavelengths in the IR range (in this work ~7-8 microns) through a difference frequency generation (DFG) process in an optical parametric amplifier (OPA, Orpheus, Light Conversion). This IR pulse is then split into two pulses (Fig. 5a-b), one of them is used to excite the sample (sample interaction) and the other is used to excite a thin Aluminum film deposited on an electron-transparent $Si_3N_4$ membrane (Fig. 5c). This membrane is positioned in the electron path, prior to the sample, and serves as the reference point of interaction in the photonic electron modulator (PELM) scheme.

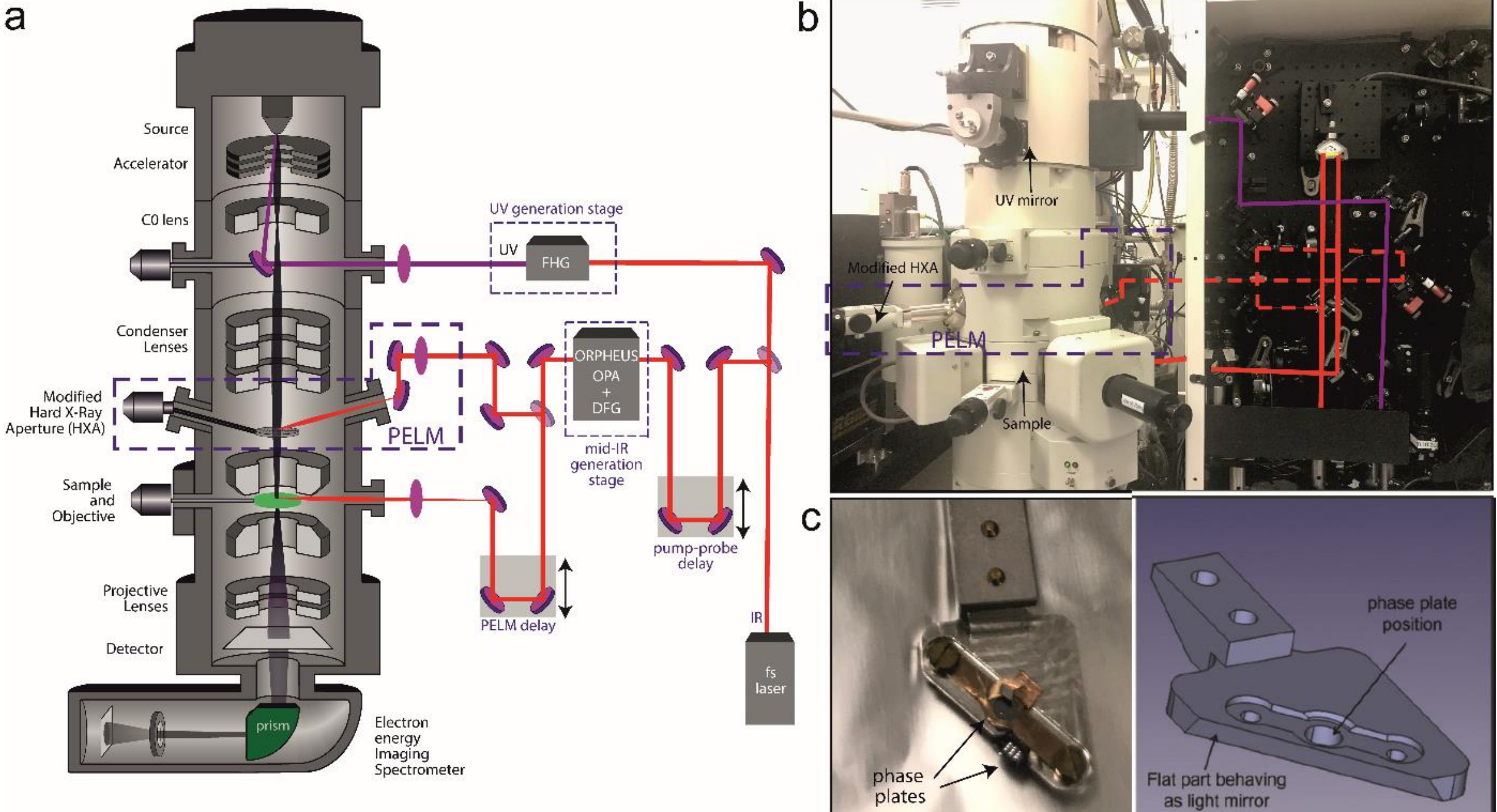

**Fig. 5 | The UTEM setup and the PELM integration.** UTEM illustration **(a)** and image **(b)** illustrating the microscope column, electron spectrometer and detectors, optical setup, and the integration of a modified Hard X-ray Aperture (HXA) at a post-condenser lens stage (PELM). The external knob of the HXA (a and b, left side) has two rigid positioning points with 5 mm lateral travel around them for positioning the reference interaction point with respect to the electron beam path. An electron-transparent thin-film sits at the place of the x-ray aperture and light enters from the optical access port on the opposite side of the column at a 20-degree angle above the horizon (dashed red line in b). Double illumination scheme (a and b, right side) implemented on the vertical board next to the UTEM. The IR laser beam is separated into two portions using a 50:50 beam splitter. One portion is guided towards the PELM (dashed red line in b), whereas the other portion is guided towards the sample (solid red line in b). **(c)** Image (left) and CAD model (right) of the modified HXA aperture connected to the platelet hosting the electron-transparent light-opaque metallic thin films for electron-light interaction. The platelet is made of Aluminum alloy, whereas the clamp is made of 0.15-mm-thick Beryllium Copper. One can observe two Si-window TEM grids (Norcada Inc.) which are coated with a 25-nm-thick Aluminum film deposited via thermal evaporation on a 10-nm-thick $Si_3N_4$ membrane. In each grid, nine slots are present to maximize the available points of interaction in case of local damage to one of the membranes. The platelet has also been cut at a specific angle allowing it to host a small metallic mirror able to reflect the light down the column towards the sample position (not used in the current work). The platelet, HXA and their integration were designed and performed in close collaboration with IDES, part of JEOL Ltd.

The temporal and spectral profiles of the IR laser pulse were characterized through an independent PINEM measurement on a metallic film, and using a grating spectrometer near the DFG stage, respectively. Noticeably, the IR pump pulse experiences some distortion along the ~5-meter optical path leading to the electron microscope column, which contributes to the observed chirp (*54*). The temporal delays between the electron pulse and the sample interaction, and between the sample and reference interactions are controlled by two motorized stages, with temporal step size of 10 fs and 1 fs, respectively, allowing sub-optical-cycle scanning of the sample and reference interaction temporal response. All delay stages are gauged to find the double time-zero, meaning that the electron is released by the UV pulse and arrives at each interaction zone simultaneously with the peak of the laser intensity. Consequently, changing the pump-probe delay in either delay stage makes the electron arrive slightly before/after the corresponding laser pulse. The delay stage on the path to the tip provides coarse control over the temporal resolution. The PELM delay stage provides fine control over the phase, i.e., the relative sub-cycle laser field interaction. The former controls the wavepacket dynamics and reveals the group velocity. The latter controls the phase velocity. In principle, the latter stage could also be shifted by a larger step to create a coarse delay and alter the wavepacket group dynamics. However, its usual fine delays maintain the wavepacket approximately unaltered due to its slower group propagation.

The TM-polarized IR pulses are focused using two lenses positioned near the microscope column, reaching a spot size of ~100 $\mu$m and average power of 4-12 mW at the sample interaction, and a spot size of ~500 $\mu$m and average power of 4-20 mW at the reference interaction. The average laser power absorbed by the sample is understood to be too low to induce significant heating of the sample, especially considering the phonon-polariton increased heat conductivity (*55*), and is below hBN's intensity damage threshold.

The sample interaction laser pulse enters the UTEM column via a side entry port, situated at the sample plane (Fig. 5a-b). At the side entry port, the laser pulse is focused via a lens and propagates perpendicular to the electron direction of motion ($z$), until reaching the sample. To prevent shadowing of the laser by the TEM grid and sample holder, our sample was tilted 35 degrees counterclockwise along the TEM holder rotation axis.

The reference interaction laser pulse enters the column above the sample interaction point, after passing through a lens, and propagates towards the Aluminum sample of the reference interaction at 20 degrees counterclockwise with respect to the lateral plane (Fig. 5a). The reference sample itself is tilted 41 degrees counterclockwise with respect to the same plane (Fig. 5c). Both tilts are done along the same axis, which is orthogonal to the laser's propagation. These tilt angles ensure that the electron and laser pulse are phase-matched on the reference interaction sample despite their different propagation velocities (a 200 keV electron travels at $0.7c$). Furthermore, the electron spot size on the reference interaction sample was intentionally kept as small as possible to ensure that the electron acquires a homogeneous phase over the entire area. The electron microscope is set such that the electron remains paraxial during its propagation from the reference to the sample, with no transverse electron dynamics. The electron beam convergence angle is less than 1 mrad, which is consistent with the assumption of paraxiality. The objective lens is turned off, there are no magnetic lenses operated between the PELM and the sample, and the condenser aperture diameter is 150 microns. Regardless of the electron imaging parameters, the transverse size of the sample (10 microns) is wavelength comparable and is much smaller than the distance between the sample and the reference. Thus, the maximal phase difference the electron can accumulate in different trajectories is on the order of $10^{-4}$ radians).

For the measurements with one point of interaction only, the Aluminum sample of the reference interaction was kept in the electron beam path, and only the laser illumination at that interaction was blocked. The average laser excitation power of the reference and signal interactions was measured individually using two power meters and was monitored routinely throughout the measurements. The lowest average laser excitation power we could detect using two points of interaction was around 2 mW (at a wavelength of 7 $\mu$m, where the signal was most visible).

After the interaction with the excited membranes (reference and sample), the free-electron energy spectrum was measured using a post-column electron energy loss spectroscopy (EELS) system with a spectrometer dispersion of 0.1 eV (Gatan Inc.). The EELS system includes a slit for producing energy-filtered transmission electron microscopy (EFTEM) images. The measurements without a reference interaction were scanned for the best slit position such that only electrons which gained energy from the sample interaction were measured. Similarly, measurements with both interactions were scanned for an optimal slit position such that only electrons gaining energy from both interactions were measured. The FWHM of the electron zero-loss peak (ZLP) without the interactions is 1.4-1.5 eV.

## 2. Free-electron Ramsey imaging (FERI): theoretical framework

At the core of PINEM lies the fundamental interaction between free-electrons and near-fields. Following typical approximations (*8,17,20,37,38*), the Hamiltonian describing the interaction between the electromagnetic field and the electron is given by $H = E_0 - i\hbar v \partial_z + evE_z(x,y,z)/\omega$, where $E_0$ is the initial electron energy, $v$ is the electron velocity, $e$ is the fundamental charge, $z$ is the electron propagation direction, $E_z(x,y,z)$ is the local electric near-field amplitude along $z$, and $\omega$ is the electromagnetic field fundamental frequency. Previous works have derived a dimensionless parameter, $g = \frac{e}{\hbar\omega}\int_{-\infty}^{\infty} dz E_z(x,y,z) e^{-iz\omega/v}$, that characterizes the interaction strength (*37,38*). In general, $g = g(x,y)$ is a complex number and a function of the transverse coordinates $(x,y)$. By measuring $g(x,y)$, one can in principle recover both the amplitude and the phase of the near-field $E_z(x,y)$, which in certain cases can suffice to reconstruct the full field information. The probability for the electron to change its energy by $l\hbar\omega$, where $l$ is an integer, is given by $P_l = |J_l(2|g|)|^2$, where $J_l$ is the l-th order Bessel function of the first kind. Furthermore, considering the electron energy spread $\Delta E$, to be a gaussian profile $G_{\Delta E}(E)$, we rewrite the above probability as:

$$P(E) = \sum_l |J_l(2|g|)|^2 \, G_{\Delta E}(E - l\hbar\omega) \,. \tag{1}$$

Applying this theory to FERI can be done by considering two points of interaction, thus we have $g = g_r + g_s$, where and $g_r$ and $g_s$ are the interaction strengths of the reference and sample fields, respectively. The FERI theory can be directly generalized for cases where the distance between the two points of interaction is large (and hence the electron pulse dispersion must be accounted for). However, this is not necessary in our setup since the distance between the two interaction points is 37 mm, negligible compared to the Talbot distance, which is on the order of a few meters in the case of mid-IR. This allows us to neglect the electron pulse dispersion between the two points of interaction. This electron then performs an effective interference between the reference and sample near-fields. The electron pulse is then sent through a dispersive magnetic prism and filtered in energy using a slit to measure only the part of the electron which gained energy in both interactions. By using Eq. 1 to build the energy filtered TEM (EFTEM) model, one will get:

$$M\big(x, y, t, \Delta\phi, E_{\text{slit}}, g_s(x, y)\big) = \int_{E_{\min}}^{E_{\max}} \sum_l G_{\Delta E}(E - \hbar\omega l) \cdot |J_l(2|g_r(\Delta\phi) + g_s(x, y, t)|)|^2 \, dE, \tag{2}$$

where $x, y$ are the sample spatial coordinates, $t$ follows the evolution of the sample dynamics, $\Delta\phi$ is the relative phase between the two laser pulses inducing the two interactions, and $E_{\text{slit}} = [E_{min}, E_{max}]$ is the electron energy-filtering slit position. Repeating this measurement while varying the sample-reference relative phase $\Delta\phi$, using sub-cycle steps, allows us to generate a complete dataset (phase scan) which fully captures the near-field dynamics at the sample.

FERI applies an algorithmic approach (*35*) to reconstruct the phase and amplitude of the field at the sample from the energy filtered measurements of the phase scan (Fig. 6), The reconstruction is based on the following FERI optimization expression:

$$\underset{|g_s(x,y)|, \angle g_s(x,y)}{\operatorname{argmin}} \sum_i \left| Y_i - M\big(x, y, t, \Delta\phi_i, E_{\text{slit}}, g_s(x, y)\big) \right|^2, \tag{3}$$

where $Y_i$ is the $i$-th measurement of the phase scan, $M$ is the measurement model (see Eq. (2)), and the reconstructed quantities are the amplitude $|g_s(x, y)|$ and phase $\angle g_s(x, y)$ of the sample field for each point in space independently.

Notably, this optimization expression is not convex, i.e., it exhibits multiple local minima, and it is well known that gradient descent methods can easily converge to a local minimum. To solve this issue, the minimization procedure scans over the relative phase $\Delta\phi$, and thanks to the low variable space (solving only for $|g_s|$, $\angle g_s$), we are able create a heatmap and perform an exhaustive search, where finding the global minimum is guaranteed (Fig. 6a3).

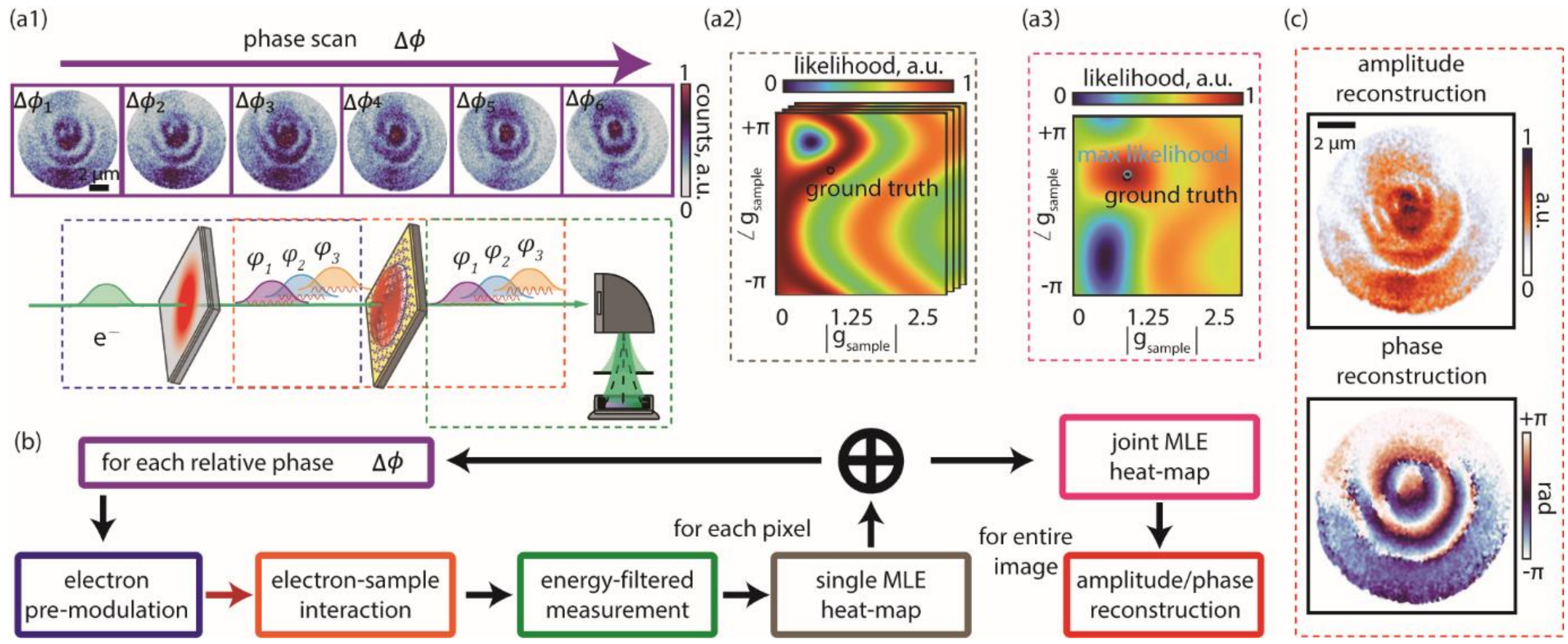

**Fig. 6 | The optimization procedure used in FERI**. Visualization **(a)** and block scheme **(b)** of the optimization process. The free-electrons are pre-modulated by a reference field before they probe the sample field and are energy-filtered to produce electron distribution measurements for each relative phase **(a1)**. For each transverse pixel $(x, y)$, several heat-maps are generated, for different relative phases between the reference and signal fields **(a2)**. The summed joint heat-map **(a3)** enables the accurate estimate of the amplitude and phase for each pixel. **(c)** The amplitude and phase reconstruction after optimizing using all the relative phases and for all the pixels.

The model in Equation (2) assumes that the reference interaction ($g_r$) is spatially uniform and that the energy filter in each position is exactly the same. However, the reference interaction has some inhomogeneity, and the automatic tuning and alignment of the Gatan Imaging Filter (GIF) is also imperfect, which is mostly noticeable when working in regimes of very weak sample interaction strength ($g_s$). We consider these non-uniformities and correct them. To this end, we measure the energy filtered image when inducing field on the reference and not on the sample (but the sample is inserted). This gives us the non-uniformities of the reference interaction due to scattering by varying thicknesses or other impurities, as well as non-uniformity due to the misalignment of the GIF. We then identify a normalization that arises from this non-uniformity and apply this normalization on the measurements.

For example, the expected reference image in Extended Fig. 1a should have been uniform in an ideal situation, however an inhomogeneity in the shape of a line across the center is apparent (the inhomogeneity is amplified in this figure for presentation clarity). To correct this inhomogeneity, we normalize the counts on the raw data at each pixel to approximate a uniform reference interaction. We carry this normalization and apply it on the PINEM measurements, as a first-order correction. The result of the reconstruction with and without these corrections are presented in Extended Fig. 1 Without the correction, some of the inhomogeneity artifacts are apparent on the reconstructed image.

### 3. Free-electron Ramsey imaging (FERI): Signal enhancement

The overall signal enhancement achieved by our FERI scheme has two contributions that each arise from a different type of coherence: the coherent nature of electron-field interactions, and the coherence between the signal and reference fields ($E_s$ and $E_r$). To explain this contribution, we recall that the conventional PINEM signal of an electron with velocity $v$ moving along $z$ depends on a dimensionless parameter $g(x,y) = \frac{q_e}{\hbar\omega}\int_{-\infty}^{\infty} dz E_z(x,y,z) e^{-iz\omega/v}$ (*37, 38*), with $q_e$ being the electron charge and $\omega$ the angular laser frequency. We denote by $g_s$ the PINEM parameter for the sample signal field $E_s$, and by $g_r$ the PINEM parameter for the reference field $E_r$.

For weak signal fields $E_s$ ($|g_s| \ll 1$) and no reference, the electron image signal scales as $|g_s|^2$. The proposed idea of coherent amplification relies on a reference field $E_r$ that is added up coherently to the signal without directly illuminating the sample. Instead, the interference is mediated *by the electron* as it is propagating between the fields. As a result, for weak fields $|g_s| \ll 1$, the electron image signal scales as $|g_s|$ (SI S3), rather than $|g_s|^2$. In this way, FERI is beneficial for situations where strong fields are difficult to apply due to weak coupling or damage to the sample. A similar enhancement factor was suggested in several works on electron interactions (*20*, *35, 57*). To the best of our knowledge, our work is the first to demonstrate the effect of free-electron coherent amplification experimentally.

The coherent amplification provided by FERI relative to conventional PINEM can be described analytically when assuming that the electron's initial energy distribution is Gaussian. When filtering only electrons with energy greater than $E_{\mathrm{filter}}$, the contrast amplification factor is given by (SI S3):

$$\frac{\sum_{n=-\infty}^{\infty}\left[J_n^2\left(2(|g_{\mathrm{r}}|)\right)-J_n^2\left(2(|g_{\mathrm{r}}|+|g_{\mathrm{s}}|)\right)\right]\cdot\mathrm{erf}\left(\frac{E_{\mathrm{filter}}-E_0-n\cdot\hbar\omega}{\sqrt{2}\Delta E}\right)}{\sum_{n=-\infty}^{\infty}\left[J_n^2(0)-J_n^2(2|g_{\mathrm{s}}|)\right]\cdot\mathrm{erf}\left(\frac{E_{\mathrm{filter}}-E_0-n\cdot\hbar\omega}{\sqrt{2}\Delta E}\right)}, \tag{4}$$

where $J_l$ are the Bessel functions of the first kind, $\Delta E$ is the electron energy spread and $E_0$ is the mean electron energy. Equation 4 shows a good qualitative agreement with our data, as shown in Fig. 4a. Expanding the amplification factor to powers of $|g_s|$ yields the $1/|g_s|$ scaling factor for weak fields, in accordance with free-electron quantum sensing protocols (*58*).

Our algorithm relies on the fact that scanning over the relative phase provides an arbitrary number of measurements for a fixed number of unknown variables. In fact, for each pixel, we need to find just two degrees-of-freedom, the amplitude, and the phase, from a scan over any number of relative phases. This type of over-constrained problem is common in interferometric reconstructions such as homodyne detection. However, unlike other interferometric techniques, the electron interaction with the field is highly nonlinear, which renders linear tools like the Radon transform generally inapplicable. The nonlinearity necessitates using the exact PINEM theory in the core of our algorithm.

We developed an optimization algorithm that solves the PINEM equations and finds the amplitude and phase for each pixel. The optimization problem is not convex (i.e., there are many local minima), so instead of gradient-descent-based methods, we create a heat-map and search for the global minima within it. Altogether, the over-constrained nature of the scheme enables extracting a good estimate of the ground truth (the correct near-field) for a wide range of parameters – as we show both experimentally and theoretically. Our algorithm is robust and applies in more general situations (e.g., regardless of bunching and of the electron energy uncertainty) and is described in detail in (*35*).

### 4. Comparison of PINEM and FERI

Extended Fig. 2 compares results of amplitude and phase reconstruction using FERI to the conventional images of PINEM, shown for different interaction strengths at the sample. The average interaction strengths over the entire image are estimated to be $|g_s| \approx [0.4, 0.8, 1.4]$ (estimated from an EELS measurement). The figure compares conventional PINEM to FERI amplitude imaging for the same electron dose on the sample per second per pixel. We now explain the comparison between the FERI images and the PINEM images. The images are acquired in a fundamentally different way, and it is thus not trivial to compare them side by side. The FERI images (Extended Fig. 2b and Fig. 1a in the main text) are achieved from a reconstruction algorithm returning a unitless constant. The PINEM images (Extended Fig. 2a and Fig. 1a in the main text) are achieved directly from the raw data measuring the number of electron-counts per pixel. For a fair comparison, we normalize the FERI and PINEM images to have their signal presented on the same axis between 0 and 1. Since the FERI method directly increases the signal range by removing the background, this approach provides a quantitative comparison of the amplification. To do this, we rescale the PINEM signal from its original arbitrary units of electron counts, to having the minimum signal at 0 and the maximum signal at $\frac{\mathrm{Max_{count}} - \mathrm{Min_{count}}}{\mathrm{Max_{count}}}$. The FERI signal is already $|g|$ in dimensionless units, so we only need to divide by $|g_{\max}|$.

One can see a clear amplification in the FERI amplitude imaging for different interaction strength, as predicted by theory (Fig. 4 of the main text). For the sake of completeness, we show that the amplification remains substantial even when further optimizing the filter parameter in PINEM while the FERI filtering is not necessarily optimized. Moreover, while conventional PINEM lacks phase information, FERI displays phase images for all the different interaction strengths. Overall, the FERI amplitude and phase imaging offer a significant improvement over conventional PINEM imaging.

**5. electron phase modulation vs. electron density modulation**

The distance between the electron modulation and the sample is known to have a critical role in determining the interaction results. Notable works on sub-cycle (phase-dependent) interactions in transmission electron microscopes aimed to have the pre-modulation stage at the exact distance that enables electron bunching (*23*, *24*). Such a scheme was recently realized by the Baum group (*24*) demonstrating the first example of sub-cycle interferometric electron imaging. However, this work did not yet show coherent amplification. The algorithm that we developed and applied for FERI can be directly extended to include electron bunching, enabling coherent amplification and extraction of amplitude and phase.

The key difference between the interferometric approach that we demonstrated here, and electron pre-bunching schemes (*24*) is the use of electron phase modulation rather than electron density modulation. Our approach (FERI) relies on electron phase modulation, without requiring a propagation distance to translate such modulations to the electron density (i.e., bunching). This difference is of particular importance for imaging lower frequencies such as the mid-IR of 2D polaritons because the bunching distance scales inversely with the frequency squared. Then for lower frequencies, the bunching distance becomes longer than the typical length available in electron microscopes. Our approach bypasses this limit and applies to any arbitrary distance between the modulation and the sample by relying on direct field interference mediated by the electron. To quantify this limit, we look at the expression of the Talbot distance (which is the characteristic length scale for the electron modulation dynamics) $z_T = \frac{4\pi m\gamma^3 v^3}{\hbar\omega^2}$. This $\frac{1}{\omega^2}$ scaling leads to the inverse relation between the bunching distance and the frequency. For $7\mu m$ modulating light and $200$keV electrons, the corresponding Talbot distance is $\sim 40m$, the relation between the Talbot distance and the bunching parameter is $|\langle b\rangle| = J_1\left(4|g_r|\sin\left(2\pi\cdot\frac{\Delta z}{z_T}\right)\right)$ with $\Delta z$ being the distance between the modulation point and the sample. In all our experiments the reference interaction strength never exceeded the value $|g_r| \approx 3$. Since the distance between the modulation and the sample interaction is $37$ mm, the maximal bunching parameter is $|\langle b\rangle| \approx 0.035$. This small bunching parameter means that only very small contrast can be achieved by relying on bunching for sub-cycle or phase imaging. Our FERI approach extracts this information without relying on bunching, making it a better fit for experiments like ours and anything in the mid-IR and lower frequencies.

## 6. Sample preparation

Samples were fabricated using viscoelastic polydimethylsiloxane (PDMS) dry transfer techniques. In the fabrication process, we mechanically exfoliate isotopically pure $h^{11}BN$ crystals grown as detailed in (*59*), on viscoelastic PDMS tapes. To reduce the amount of chemical residue and to conserve the hBN crystal, no tape was used in the exfoliation. Rather, exfoliation was performed directly with low retention PDMS (commercially available from GELPAK, X0 retention, in DGL or PF format). After the initial exfoliation steps, the PDMS was visibly covered with many hBN crystals, at which point a fresh PDMS sheet was used to pick up (and exfoliate further) a portion of the hBN. Further exfoliation using fresh pieces of PDMS was repeated until the typical flake thickness was estimated, based on optical microscope examination, to be close to the target thickness.

After 2-5 such rounds, a last round of exfoliation was performed directly on the stamp, by attaching and separating it to one of the hBN covered PDMS sheets. By controlling the speed of the exfoliation, especially in this last step, we change the prevalent physical process; at slower speeds, flakes tend to move from one PDMS to another and at higher speeds they tend to exfoliate, but at the risk that they apply strain and crack or break the flakes. To better estimate the thickness of the hBN flakes we performed EELS log-ratio measurements and obtained 40-50 nm for the investigated flake.

The target on which the hBN was dropped was a 2-by-2 array of 10 µm circular holes (Extended Fig. 3) in a 50 nm-thick $Si_3N_4$ membrane (available commercially from Norcada Inc.), upon which a layer of Au (with a Ti seed layer) was thermally evaporated. Nominal thickness of the Au (Ti) is 10 nm (2 nm), determined by AFM measurements on a calibration chip.

To drop the flakes, the membrane was heated to 60 ℃, after which the PDMS was slowly brought into contact with the membrane window. We constantly tracked the locations of the flake, membrane window, and contact front between the PDMS and the membrane's support substrate under an optical microscope. The PDMS was then lifted slowly, leaving the hBN attached to the Au layer. In some cases, if the flake appeared not to connect to the membrane, the temperature was raised up to 90 ℃ before the stamp was lifted. The regions of interest in the experiment were the pre-etched holes in the membrane. At these regions the hBN flake is free-standing, thereby minimizing electron losses in the material and reducing substrate-related losses of the phonon-polariton. Our experimental studies focused on one of these windows, indicated in Extended Fig. 3, where the quality of the particular hBN flake was superior. The other holes showed related polaritonic behavior, but with weaker laser coupling strength and less pronounced features. Multiple measurements with the same settings showed little variation in the acquired signal, indicating that the system is reproducible and that the flakes did not deteriorate over time.

### 7. Correcting for 3D sample tilt using image processing

For quantitative analysis of the phonon-polariton properties, such as the phase velocity, one must first correct the sample tilt in space with respect to the propagation direction of the free-electrons. Due to the two-axes tilt, the raw data exhibits an oval instead of a circular shape.

Extended Fig. 4 presents the procedure that corrects for the tilt and retrieves the original circular shape.

## Data Availability

All the key data that support the findings of this study are included in the Article and its Supplementary Information. Further datasets and raw measurements are available from the corresponding authors upon reasonable request.

## Methods-only references

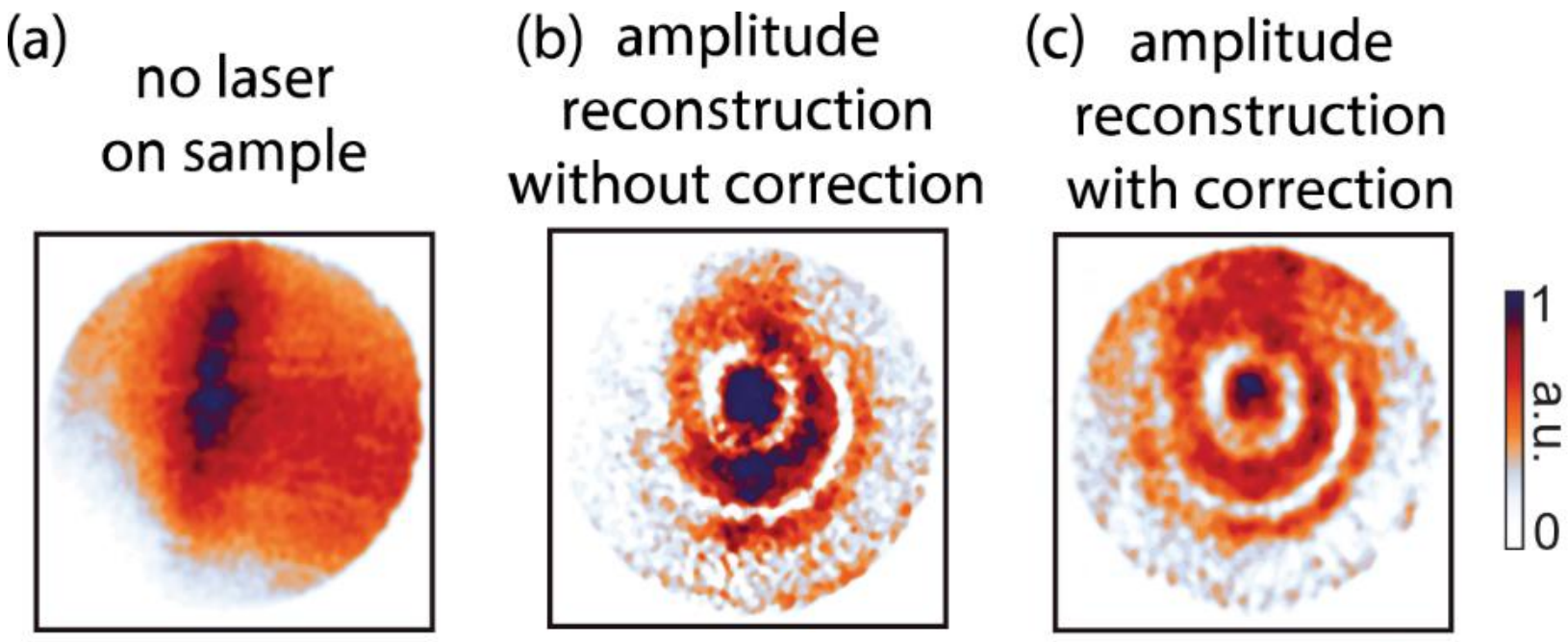

**Extended Fig. 1 | Reference non-uniformity correction**. **(a)** Energy filtered measurement with light induced only on the reference interaction. **(b,c)** Amplitude reconstruction for weak $g_s$ without **(b)** and with **(c)** correction**.** All the images are normalized to be between 0 and 1 such that the features are visible.

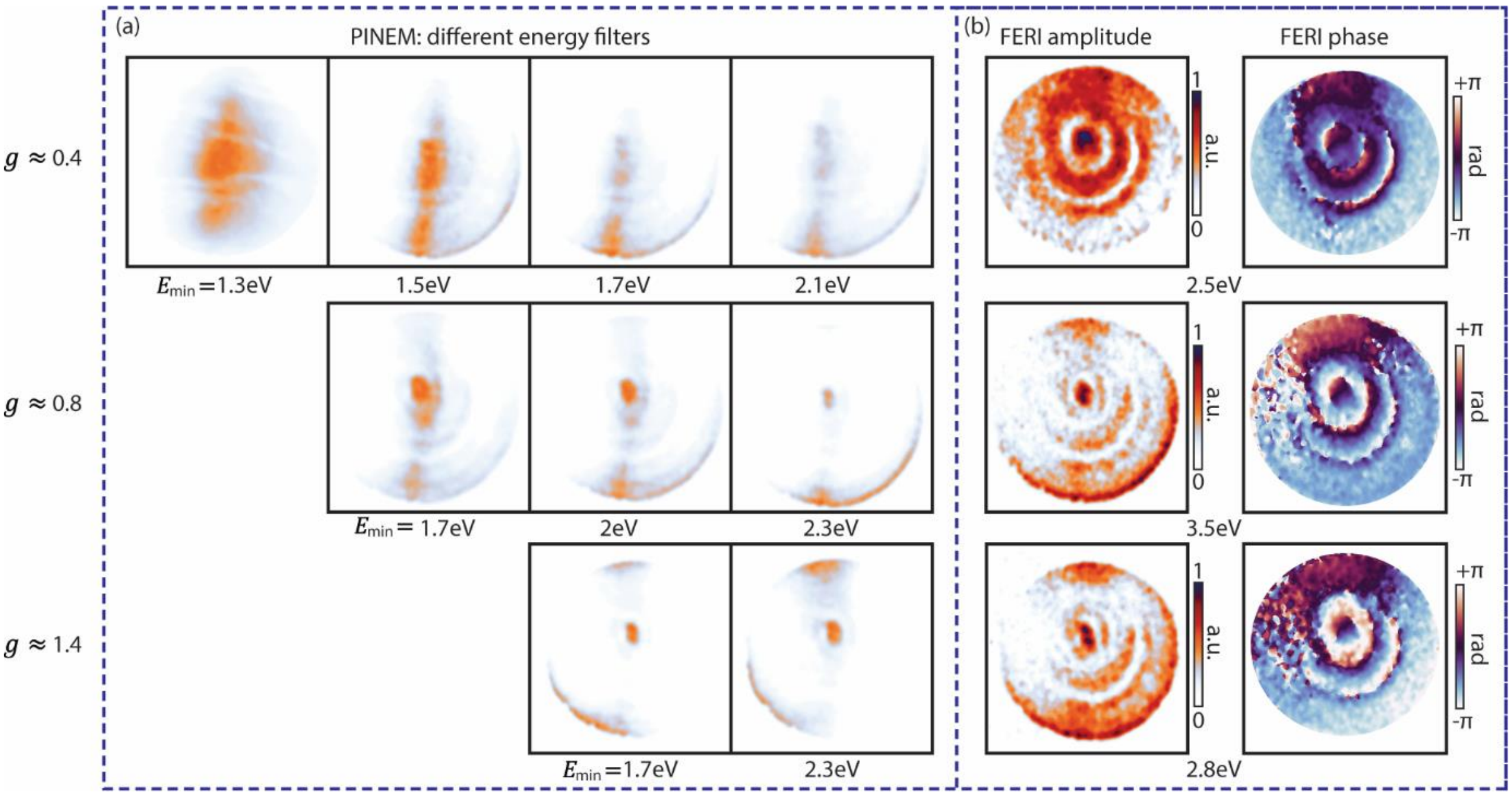

**Extended Fig. 2 | Phonon-polariton amplitude and phase reconstruction for different illumination intensities compared to conventional PINEM imaging.** Data acquired using three different average interaction strengths over the entire image: $|g_s| \approx 0.4, 0.8, 1.4$, top to bottom, respectively. **(a)** conventional PINEM amplitude imaging for the three different interaction strengths and for different energy filtering (10 eV slit width, cutoff energy is marked at the bottom of each image), acquired for the same time-delay. **(b)** FERI amplitude imaging for the same electron dose on the sample. As predicted by theory (Fig. 4 of the main text), the largest amplification is seen for the weakest interaction strength. The right column shows FERI phase images for the different interaction strengths.

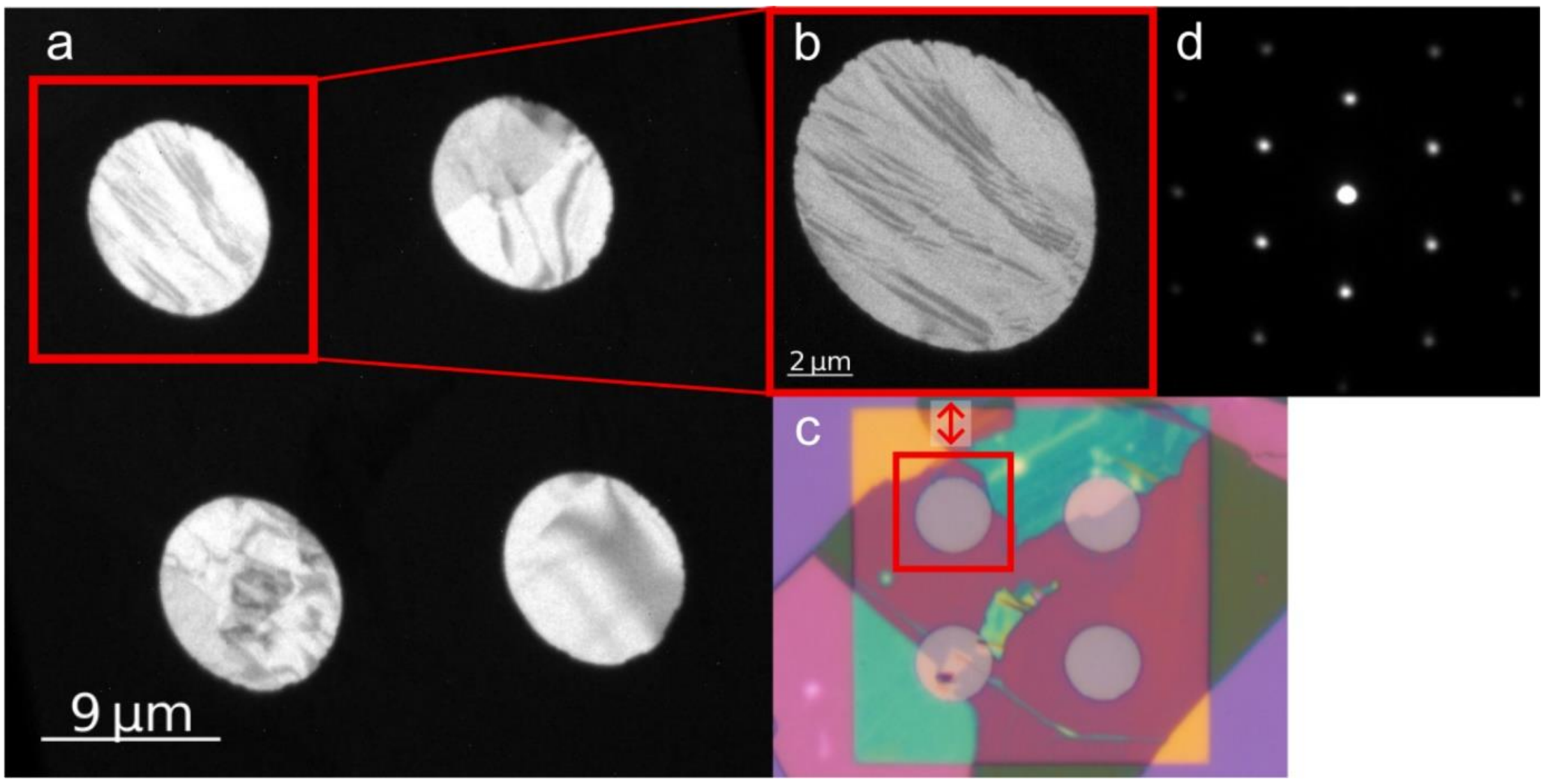

**Extended Fig. 3 | Sample images. (a,b)** TEM images of the investigated hBN flake. The experimental results shown in the main text could be reproduced in different windows of the flake, however, the highlighted window (b) gave the best coupling efficiency and thus strongest signal. **(c)** Electron diffraction pattern of the hBN flake highlighted in (b), confirming that the whole flake is mono-crystalline. **(d)** Optical microscope image of the investigated hBN flake. Different colors correspond to different hBN thicknesses and are formed due to optical interference in the hBN and substrate. The investigated window shows a uniform thickness.

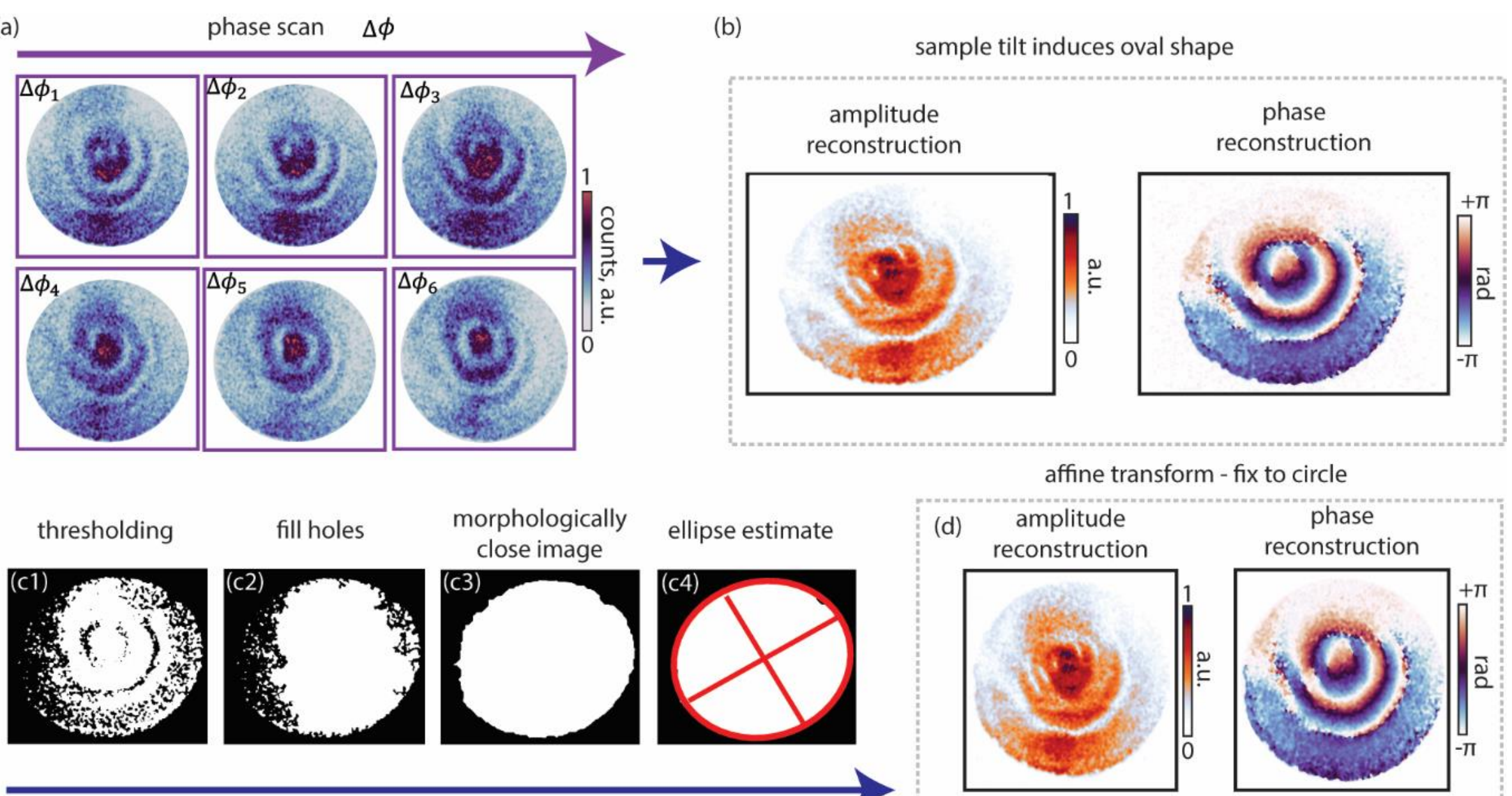

**Extended Fig. 4 | Image post-processing revealing the phonon-polaritons properties. (a)** The relative phase scan produces the FERI tilted raw measurements which are used to reconstruct the amplitude and phase **(b)**. **(c1-4)** The process for the ellipse estimation: **(c1)** convert the amplitude reconstructed image to black and white by thresholding. **(c2-3)** perform morphological image processing (56). **(c4)** estimate the ellipse equation through connected component analysis. **(d)** Perform an affine transformation on the reconstructed images using the inverse of the estimated ellipse equation.